\documentclass[english]{article}
\usepackage{authblk}
\usepackage[T1]{fontenc}
\usepackage[utf8]{inputenc}
\usepackage{geometry}
\usepackage{color}
\usepackage{amsmath}
\usepackage{cite}
\usepackage{amssymb}
\usepackage{graphicx}
\usepackage{babel}
\usepackage{hyperref}
\usepackage{pdfpages}
\usepackage{soul}
\usepackage{makecell}
\usepackage{babel}

\usepackage[normalem]{ulem}
\usepackage{cancel, xstring}
\usepackage{xparse}

\NewDocumentCommand{\delete}{m}{
  \IfBeginWith{\detokenize{#1}}{\detokenize{\eqref}}{
    \sout{\text{#1}} 
  }{
    \ifmmode
      \text{\sout{\ensuremath{#1}}}
    \else
      \sout{#1}
    \fi
  }
}

\newcommand{\add}[1]{{#1}}

\newcommand{\remove}[1]{}

\title{Inverse Design of Parameter-Controlled Disclination Paths}
\author[1]{Yehonatan Tsubery}
\author[1]{Hillel Aharoni}
\date{}
\affil[1]{Department of Complex Systems, Weizmann Institute of Science, Rehovot 7610001, Israel}

\begin{document}
\maketitle

\begin{abstract}

Topological defects, such as disclination lines in nematic liquid crystals, are fundamental to many physical systems and applications. In this work, we study the behavior of nematic disclinations in thin parallel-plate geometries with strong patterned planar anchoring. Building on prior models, we solve both the forward problem—predicting disclination trajectories from given surface patterns—and an extended inverse problem—designing surface patterns to produce a tunable family of disclination curves under varying system parameters. We present an explicit calculation for pattern construction, analyze parameter limitations and stability constraints, and highlight experimental and technological applications.
\end{abstract}

\section*{Introduction}

Topological defects play a crucial role in various physical systems, from condensed matter physics (vortices in superconductors, dislocations in crystals) to cosmology (cosmic strings, monopoles). Many physical phenomena in these systems strongly depend on defect characteristics and interactions, including their formation and combination rules, charges, geometry, forces, etc. Therefore, understanding and controlling defect structure and behavior \emph{in situ} may be extremely useful for a variety of applications.

Disclination lines in nematic liquid crystals (NLCs) are a simple and elegant class of topological defects, making them a good subject for both theoretical and experimental studies \cite{deGennesBook}. They are one-dimensional topologically-protected singularities in the nematic orientational order \cite{mermin_topological_1979}, and may appear either as transients or as a result of incompatible boundary conditions or elastic frustration. Nematic disclination lines have applications in directed assembly of particles and molecules \cite{musevic2006two,blanc_ordering_2013,guo2024directing,wang_topological_2016,luo_tunable_2018}, optical devices \cite{brasselet_tunable_2012,meng_topological_2023}, microfluidics \cite{sengupta_topological_2013} and others. With major recent advances in spatial patterning of liquid crystal alignment \cite{nys2020patterned,modin2024spatial}, NLC disclinations make excellent candidates for designing and tuning defect geometries.

\begin{figure}[ht]
\centerline{\includegraphics[width=0.9\columnwidth]{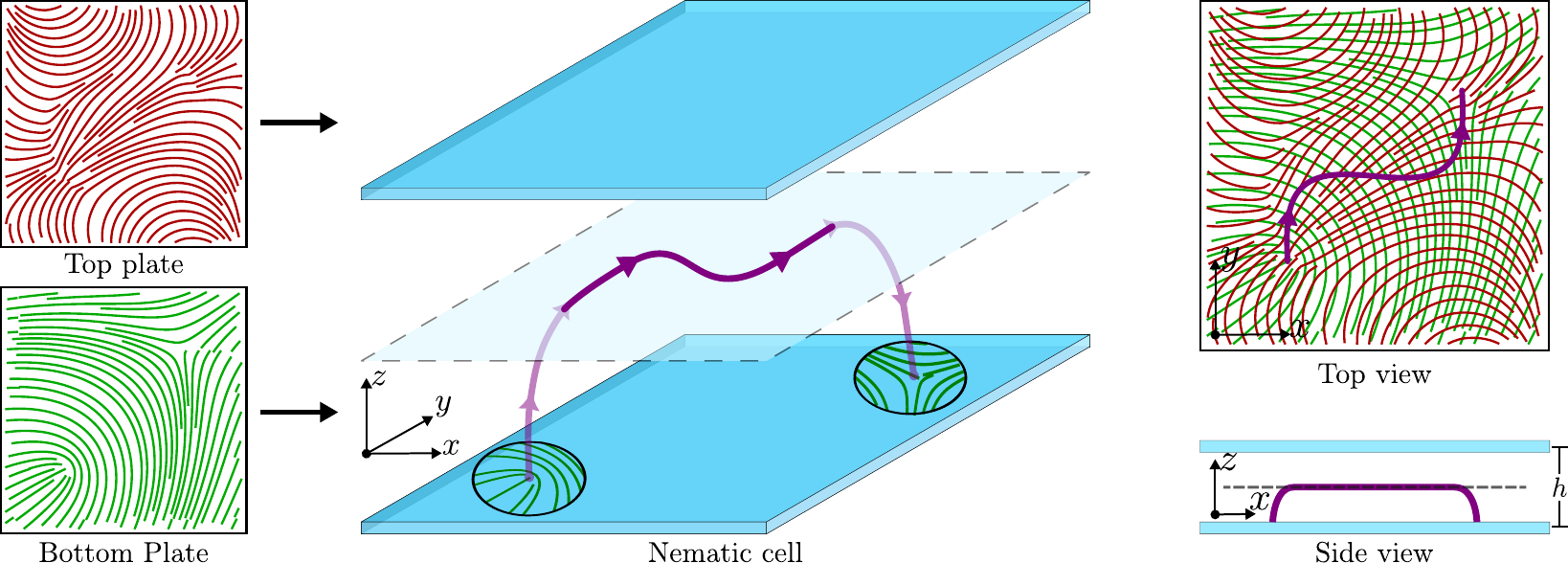}}
\caption{Sketch of the parallel plate setup. An NLC is placed between two parallel patterned thinly spaced plates, with strong planar anchoring. Half-integer disclinations in the 2D surface pattern are endpoints to disclination lines. At equilibrium, these lines traverse horizontally along the mid-plane between the two boundaries (except near their ends). Their top-view projection is discussed in this paper.}
\label{fig1}
\end{figure}

In the \emph{parallel-plate setup}, a thin layer of NLC  is placed between two flat plates (typically glass). The plates are pretreated using a variety of chemical, optical and mechanical techniques \cite{modin2024spatial,geary1987mechanism,wen2002nematic,kim2002nano, newsome1998laser, chigrinov_photoalignment_2008} to impose a molecular orientation in a preferred direction within the plane, sufficiently strong to constrain the nematic director in their immediate vicinity. These conditions are known as \emph{strong patterned planar anchoring} and are abundant in experiment. Disclination lines in such systems are induced either via surface defects or via strong twist reversals \cite{sunami_shape_2018}. The parallel-plate setup (Fig.~\ref{fig1}) enables realization of line arrays \cite{chuckArbitraryArray, guo_photopatterned_2021, nys_nematic_2022,harkai_electric_2020,moirewang2024}, arbitrary defect geometries \cite{modin2023tunable, sunami_shape_2018}, and defect line sources \cite{frankread}.

For sufficiently thin cells, the director field remains planar not only on the boundaries but also in the entire bulk \cite{modin2023tunable}. This observation leads to a half-integer classification of line defects, similar to that of point disclinations in 2D \cite{mermin_topological_1979, long2021geometry}. In \cite{modin2023tunable}, equilibrium states of disclination lines were studied based on this assumption of bulk planarity. A closed-form equation for disclination trajectories was derived, taking into account the strong anchoring conditions, nematic elasticity, and line tension. Furthermore, \cite{modin2023tunable} resolved the \emph{inverse problem} of designing a cell in which a specific desired curve is in equilibrium. A different inverse design approach was introduced in \cite{sunami_shape_2018}, based on designing very strong elastic potential wells to overcome line tension.

Building on the framework in \cite{modin2023tunable}, we first address the forward problem: given prescribed surface patterns on two parallel plates, we determine the resulting disclination path in equilibrium. For the inverse problem, we observe that fully controlling the anchoring on the entire boundary surface to realize just a single curve shape leaves a multitude of unused degrees of freedom. This motivates an \emph{extended inverse problem}: designing the confining surface patterns such that \emph{a family of curves} is obtained when changing system parameters. We show how to explicitly calculate such patterns, discuss the limitations of different parameter types, and explore potential applications.

\section*{Results}

We consider an NLC in the parallel plane setup with strong patterned planar anchoring. The surface patterns for the top and bottom plates are denoted $\theta_t(x,y)$ and $\theta_b(x,y)$, respectively, representing planar director angles (defined modulo $\pi$). It is assumed that the thickness of the cell $h$ is much smaller than all lateral length scales in the problem. We assume the standard Frank free energy density for elastic distortions,
\begin{equation}\label{Frank}
    \mathcal{F}_\textrm{el} = \tfrac{1}{2}K_1(\nabla\cdot\hat{\boldsymbol{n}})^2+\tfrac{1}{2}K_2(\hat{\boldsymbol{n}}\cdot\nabla\times\hat{\boldsymbol{n}})^2+\tfrac{1}{2}K_3\|\hat{\boldsymbol{n}}\times\nabla\times\hat{\boldsymbol{n}}\|^2,
\end{equation}
in the two constant approximation $K_3=K_1$. In addition we assume constant disclination line tension $\gamma$ (see \cite{modin2023tunable}).

In \cite{modin2023tunable}, the equilibrium shape of disclination lines in the above model is calculated using the following steps:
\begin{enumerate}
    \item For a given disclination line shape \add{$\Gamma$}\remove{$\boldsymbol{\Gamma}$}, the equilibrium state of the director field in the bulk is calculated.
    \item The force $\boldsymbol{f}(\boldsymbol{\Gamma})$ acting on the line is calculated using analogy with magnetostatics.
    \item Equilibrium shapes, for which $\boldsymbol{f}(\boldsymbol{\Gamma})=0$, are found.
\end{enumerate}
It is shown that disclination lines connect surface defects according to connectivity rules of two-dimensional nematics. These lines traverse laterally along the mid-plane between the two boundaries, except near their ends. \add{Consequently, lateral disclinations in the planar regime are locally of the twist type \cite{long2021geometry}}. This renders the entire problem two-dimensional (Fig.~\ref{fig1}). The in-plane force acting on a lateral line segment is
\begin{equation}\label{force}
\boldsymbol{f}=\gamma\boldsymbol{\kappa}+\frac{2\pi qK}{h}\overline{\Delta\theta}\left(\hat{\boldsymbol{z}}\times\hat{\boldsymbol{T}}\right),
\end{equation}
where the system parameters are $h$ the cell thickness, $\gamma$ the line tension, and $K\equiv\sqrt{K_1K_2}$ the elastic constant. The geometric measurables are $\hat{\boldsymbol{T}}$ and $\boldsymbol{\kappa}$, the Frenet-Serret tangent and curvature vector of the disclination curve, respectively, and $q$ the disclination charge. Note that $\hat{\boldsymbol{T}}$ and $q$ are defined ambiguously by an arbitrary choice of direction, however the product $q \hat{\boldsymbol{T}}$ and therefore eq.~\eqref{force} remain invariant. $\overline{\Delta\theta}\cong \theta_t-\theta_b\pm q\pi~(\text{mod } \pi)$ is the total rotation of the director between the top and bottom plates. Unlike \cite{modin2023tunable}, it is averaged over the two sides of the disclination line to remove ambiguity in the definition.
We simplify Eq.~\eqref{force} by defining $\hat{\boldsymbol{T}}_\perp\equiv\hat{\boldsymbol{z}}\times\hat{\boldsymbol{T}}$ and setting $\widetilde{\Delta\theta}=2\pi q\overline{\Delta\theta}$, $\kappa=\boldsymbol{\kappa}\cdot\hat{\boldsymbol{T}}_\perp$ and $\lambda=\frac{\gamma}{K}h$. As seen in \cite{modin2023tunable}, $\lambda$ is the line-tension-induced smoothing scale of the disclination curve. 
We now obtain the simple form
\begin{equation}\label{simpleforce}
    \boldsymbol{f}=f\,\tfrac{K}{h}\hat{\boldsymbol{T}}_\perp \qquad\text{where}\qquad f=\lambda\kappa+\widetilde{\Delta\theta}.    
\end{equation}
Equilibrium disclination lines are then obtained by a stable force balance
\begin{subequations}\label{stableforcebalance}
    \begin{gather}
        \lambda\kappa+\widetilde{\Delta\theta} = f = 0\label{forcebalance},\\
        \frac{\delta f}{\delta \Gamma} < 0,\label{stableforce}
    \end{gather}
\end{subequations}
where $\delta{\boldsymbol{\Gamma}}=\delta \Gamma\,\hat{\boldsymbol{T}}_\perp$. \add{Eq.~\eqref{stableforce} states that for an equilibrium path to be stable, an infinitesimal variation, $\delta{\boldsymbol{\Gamma}}$, must induce a restoring force.}

\subsection*{Equilibrium disclination paths}

In experimental systems such as \cite{modin2023tunable,sunami_shape_2018,nys_nematic_2022} the patterning on the top and bottom plates is fixed. Eq.~\eqref{forcebalance} is then a local second-order system of ordinary differential equations (SODE or \emph{spray}) \cite{shenSpray} for disclination paths:
\begin{equation}\label{spray}
    \ddot{\boldsymbol{\Gamma}} = (\ddot{x},\ddot{y}) = -\frac{\widetilde{\Delta\theta}}{\lambda }(-\dot{y},\dot{x})
\end{equation}
where $\boldsymbol{\Gamma}(s)=\left(x(s),y(s)\right)$ is the arclength-parametrized disclination path \add{and with $\dot{\boldsymbol{\Gamma}}=\frac{d\boldsymbol\Gamma}{ds}$}. Eq.~\eqref{spray} uniquely determines an equilibrium path (a \emph{geodesic} of the spray) given an initial condition, namely its position and direction at a point.

In physical scenarios, a full set of initial conditions at a single point is rarely the case. Rather, we often find boundary conditions that are fixed (e.g. pinning to surface defects), free (e.g. end of nematic region) or mixed (e.g. end of patterned surface region). In such cases uniqueness of the solution to eq.~\eqref{spray} is not a priori guaranteed, however could be derived locally by means of the shooting method or similar tools. Many systems \cite{nys_nematic_2022,jiang2022active,harkai_electric_2020,guo_photopatterned_2021} indeed show multi-stability of solutions.

Equations \eqref{stableforcebalance} can be derived using calculus of variations from a simple free energy functional (see Appendix A). In addition to eq.~\eqref{forcebalance} as its Euler-Lagrange equation, such analysis gives the exact form of the boundary term; disclination lines that end on a free boundary must do so perpendicularly to the boundary. Additionally, it allows addressing the second variation to verify the stability of disclination paths, which we use in the following.

\subsection*{Inverse problem}

After solving equilibrium paths given surface patterns, we naturally turn to the inverse problem. Given a plane curve $\Gamma$, can we design a cell with suitably chosen surface patterns $\theta_{t}$
and $\theta_{b}$, such that disclination line will emerge in the shape of $\Gamma$? This question can be answered promptly by reading eq.~\eqref{forcebalance} backwards. For a given curve $\Gamma$, $\kappa$ is known and eq.~\eqref{forcebalance} can simply be solved algebraically for $\overline{\Delta\theta}$, thus for $\theta_{t}$
and $\theta_{b}$. This algorithm is not only simple but also very much degenerate, since $\overline{\Delta\theta}$ is determined only on the curve $\Gamma$ itself. Therefore, multiple surface patterns may give rise to the same given curve $\Gamma$ \cite{sunami_shape_2018,modin2023tunable}.

\begin{figure}[ht]
\centerline{\includegraphics[width=0.9\columnwidth]{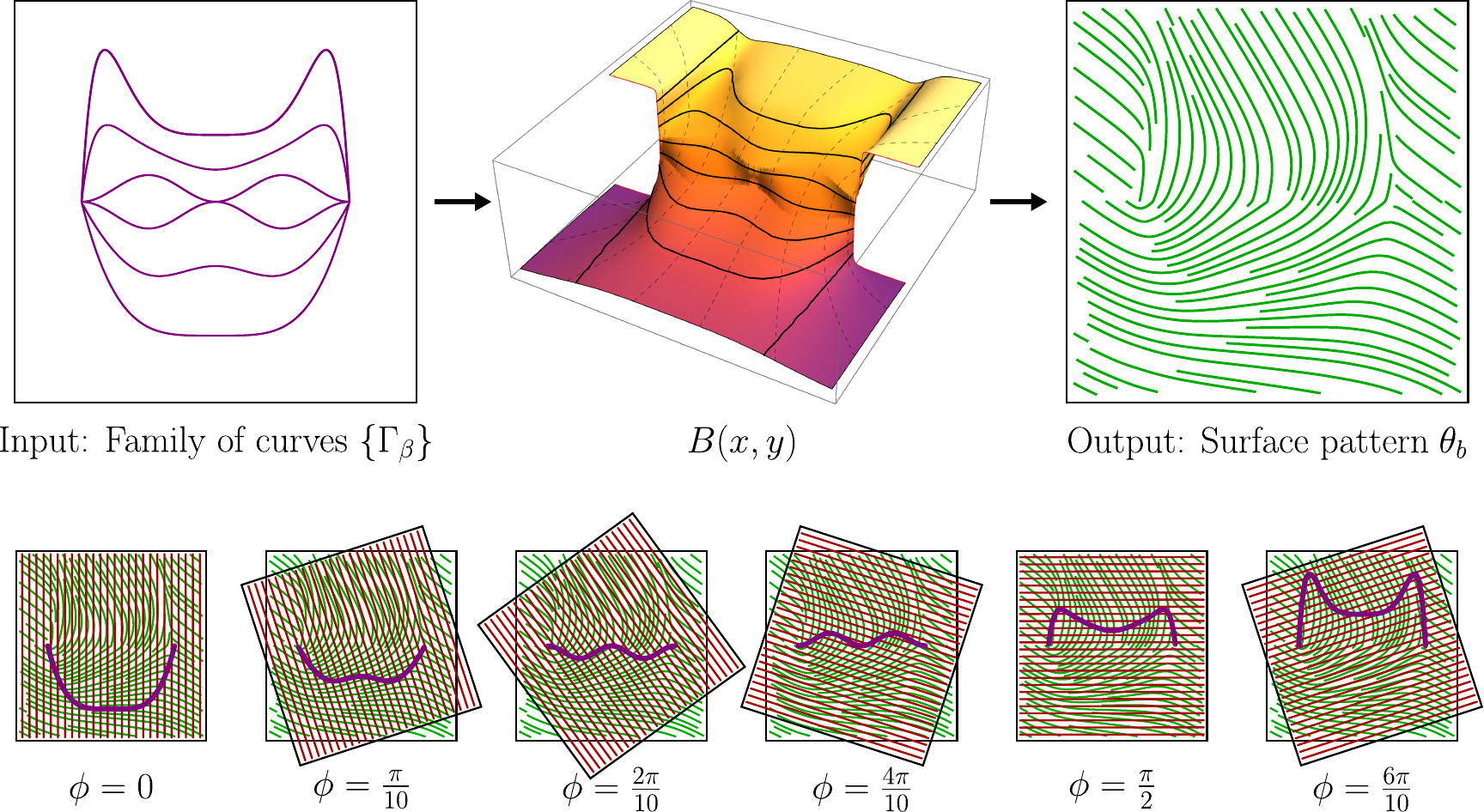}}
\caption{Inverse design workflow. A desired family of curves $\left\{ \Gamma_{\beta}\right\}$ is given as input. A related function $B(x,y)$ is constructed in agreement with eq.~\eqref{Bdef}, from which we calculate surface patterns \add{(Dataset S1)} for the confining plates (the calculation depends on the choice of physical control parameter $\beta$, see text\add{, Table~\ref{tab:my_label}} and figure.~\ref{fig3}). Bottom: at each value of $\beta$ (here the rotation $\phi$ of the top plate), $\Gamma_{\beta}$ emerges as the equilibrium disclination curve.
}
\label{fig2}
\end{figure}

We thus turn to describe a system whose force balance equations eq.~\eqref{stableforcebalance} depend on a parameter $\beta$. For each value of $\beta$, we obtain a unique spray (eq.~\eqref{spray}) and its associated geodesics. Now, given a set of plane curves $\left\{ \Gamma_{\beta}\right\}$, we aim to find a single pair of surface patterns $\theta_{t}$ and $\theta_{b}$, that will realize each curve $\Gamma_\beta$ at the corresponding value of the parameter $\beta$.
We represent the input family of desired plane curves with an auxiliary function $B(x,y)$, whose level sets are the target curves:
\begin{equation}\label{Bdef}
\Gamma_{\beta}=\left\{ (x,y)\in\mathbb{R}^{2}\,|\,B(x,y)=\beta\right\}.
\end{equation}
This notation implicitly assumes that $\Gamma_\beta$ that belong to different $\beta$ are mutually non-intersecting (except for isolated singular points, which we handle later). In many cases this assumption is not necessary, and intersecting curves can be achieved for different values of $\beta$, however since it makes the formalism simpler we assume it for clarity. We now turn to write equations~\eqref{stableforcebalance} in terms of the input function $B(x,y)$ and (algebraically) solve them for $\theta_{t/b}$ (Fig.~\ref{fig2}).

\remove{
\paragraph{Force balance.}}
We interpret $\beta$ as some physical parameter which alters
the force equation. \add{$\beta$ can represent either a system parameter, e.g. the temperature, or a relative 2D transformation between the two plates, e.g. rotation or translation. $\beta$ affects the force balance equation by either of these two mechanisms, or both.}\remove{ Changing $\beta$ can represent changes in the system parameters, e.g. the temperature, or a relative 2D rigid transformation between the two plates which we denote by ${\bf L}_{\beta}$.}
Since for every $\beta$ the path \add{$\Gamma_{\beta}$}\remove{$\boldsymbol{\Gamma}_{\beta}$} is at equilibrium, we write eq.~\eqref{forcebalance} for all values of $\beta$:
\begin{equation}\label{betaforcebalance}
\add{\forall\beta,\quad}
\lambda(\beta)\kappa(\Gamma_\beta)+\widetilde{\Delta\theta}(\beta,\remove{{\bf L}_{\beta}}\Gamma_\beta)=0.
\end{equation}

\add{
        We now use eq.~\eqref{betaforcebalance} to rewrite eqs.~\eqref{stableforcebalance} in terms of $B(x,y)$ to explicitly extract the balance at each $x,y$ position:
        \begin{subequations}\label{Bstableforcebalance}
        \begin{gather}
            \widetilde{\Delta\theta}(B,(x,y)) =-\lambda(B)\kappa(x,y)\label{Bforcebalance}\\
            \frac{\partial f}{\partial \beta}(\hat{\boldsymbol{T}}_\perp\cdot\nabla B) >0.\label{Bstableforce}
        \end{gather}
        \end{subequations}
        
    Since disclination lines are level sets of the function graph $B(x,y)$, their in-plane curvature $\kappa(x,y)$ can be explicitly written in terms of the function $B(x,y)$ and its derivatives:
        \begin{equation}\label{kappa}
           \kappa(x,y)=\sigma \frac{B_{x}^{2}B_{yy}-2B_{x}B_{y}B_{xy}+B_{y}^{2}B_{xx}}{(B_{x}^{2}+B_{y}^{2})^{3/2}},
        \end{equation}
    where subscripts indicate partial derivatives and $\sigma = -\text{sign}\left(\hat{\boldsymbol{T}}_\perp\cdot\nabla B\right)$.

    Equations~\eqref{Bstableforcebalance} can be readily used to convert a desired family of curves as a function of a control parameter into patterns that will realize it. Eq.~\eqref{Bforcebalance} spells out the pattern, and eq~\eqref{Bstableforce} asserts its stability or lackthereof. Eq.~\eqref{Bstableforce} also sheds light on the physical parameter-dependent behavior of curves for different control parameters, as we will see next.
    }

\subsection*{Control Parameters}
\begin{figure}[ht]
\centerline{\includegraphics[width=1\columnwidth]{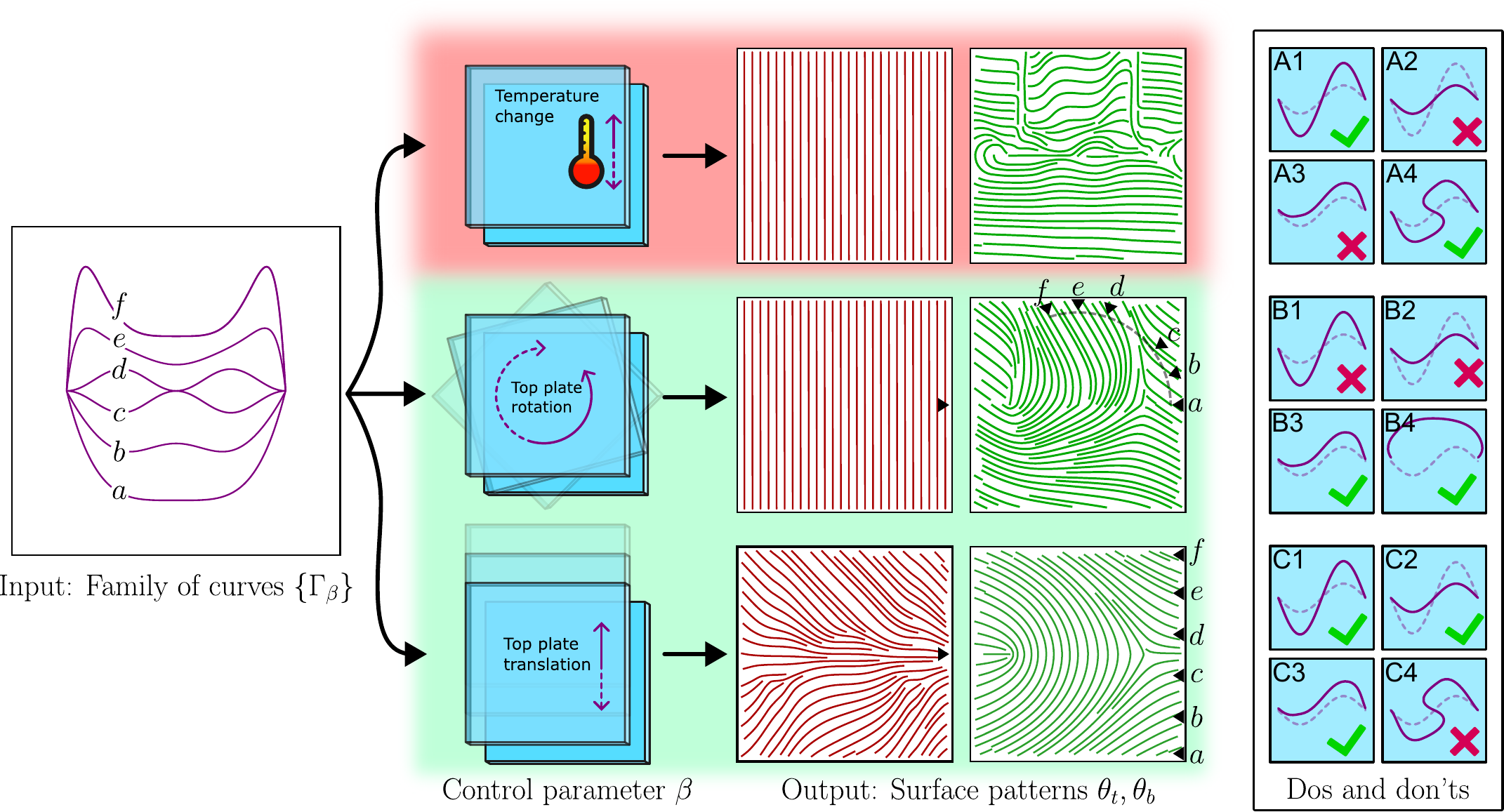}}
\caption{Control parameters. Given an input family of curves \add{(subset labeled a-f)}, and depending on the available control parameter -- temperature (top), plate rotation (middle), or plate translation (bottom) -- the top and bottom surface patterns are calculated \add{(Datasets S3, S1 and S3-4, respectively)} \add{using equations presented in Table~\ref{tab:my_label}}\remove{(eqs.~\eqref{theta1}, \eqref{theta2}, and \eqref{theta3}, respectively)}. Background color expresses meeting the stability criteria for each scenario \add{(Table~\ref{tab:my_label})}\remove{(eqs.~\eqref{Stability1B}, \eqref{Stability2}, and \eqref{Stability3}, respectively)}.
\add{To achieve a specific curve, align the triangle indicator on the top plate with the corresponding labeled indicator on the bottom plate through rotation or translation.}
\add{The rightmost column presents archetypes of allowed and forbidden curve transitions for each control parameter.}
\remove{Systems with rotating or translating plates can realize the cat curves in this example, however a system with only changing temperature cannot.}
\add{The cat-shaped curves in this example can be stably realized in systems with rotating or translating plates, however not in a system with only changing temperature.}}
\label{fig3}
\end{figure}

\remove{
    We turn to demonstrate explicit formulae for patterning cells that give rise to desired families of curves, for different physical control parameters $\beta$ that are of common use in the literature (Fig.~\ref{fig3}).}
\add{We turn to apply equations~\eqref{Bstableforcebalance} to different physical control parameters $\beta$ that are of common use in the literature (Fig.~\ref{fig3}). In each case, we calculate explicit ready-to-use cell patterns that give rise to desired families of curves (see Datasets S1-5). The resulting pattern formulae are summarized in Table~\ref{tab:my_label}.}

\add{
        \begin{itemize}
        \item 
        \textbf{Temperature.}
        We choose the control parameter $\beta$ to be the smoothing scale $\lambda$ associated with line tension. This parameter can be controlled via the temperature -- see \cite{modin2023tunable} (one may also choose $\beta$ to be the cell thickness $h$, which is mathematically equivalent). Intuitively, we can think of a disclination as a tensed string. Increasing the parameter $\lambda$ effectively increases the tension by pulling on the string's ends. With this picture in mind, stability suggests that cooling (increasing $\lambda$) shifts the equilibrium of the string inwards with respect to its own curvature, ``ironing out'' the curves. Conversely, heating (decreasing $\lambda$) shifts the string outwards. Therefore, transitions result in local length shortening with an increase in $\lambda$ and local length elongation when $\lambda$ is decreased (fig.~\ref{fig3}).
        In this case, as seen in Table~\ref{tab:my_label}, only $\overline{\Delta\theta}\cong\theta_t-\theta_b\pm q\pi~(\text{mod}~\pi)$ is determined and there remains gauge freedom to split it between $\theta_t$ and $\theta_b$. Convenient gauges are $\theta_t=\tfrac{\overline{\Delta\theta}}{2}$ and $\theta_b=q\pi-\tfrac{\overline{\Delta\theta}}{2}$; or, $\theta_{t}=q\pi+\overline{\Delta\theta}$ and $\theta_b=0$; or, $\theta_b$ contains all surface defects and is otherwise harmonic and $\theta_t=\overline{\Delta\theta}+\theta_b$.

        \item 
        \textbf{Plate rotation.}
        We consider a simple setup, similar to \cite{frankread,zhang2025nonreciprocal}, where the top plate is at a constant angle and is being rotated with respect to the bottom plate. The control parameter $\beta = \phi$ in this case is the top plate rotation angle. When rotating the top plate counter-clockwise we are uniformly adding positive twist to the NLC between the plates.
        As a result, disclination lines, at which there is a discontinuity in the total twist between the plates, are shifted depending on their charge $q\hat{\boldsymbol{T}}$.
        A simple analog would be thinking of the disclination as a one-dimensional membrane in a non-uniform pressure field. Rotation is equivalent to a non-conservative force acting to rotate the fluid, resulting in a uniform pressure difference between the two sides of the membrane.
        A closed disclination loop is equivalent to a balloon being inflated or deflated upon rotation.
        Back to our setup, stability implies that a curve can deform only towards one of its sides as the top plate is rotated. This direction is $q\hat{\boldsymbol{T}}_\perp$ for CCW rotation and $-q\hat{\boldsymbol{T}}_\perp$ for CW rotation. Equivalently, at its ends the curve must co-rotate with the top plate around a positively charged surface defect on the bottom plate, and counter-rotate with the top plate around a negatively charged defect.

        \item
        \textbf{Plate translation.}
        We consider a translation of a non-uniform top plate in direction $\hat x$ by distance $\beta = X$. Unlike the previous two examples, eq.~\eqref{Bforcebalance} is now truly non-local; this fact adds a step when calculating the top and bottom patterns as seen in Table~\ref{tab:my_label}. 
        This case is somewhat similar to plate rotation, since at each point there is an induced top-pattern rotation proportional to $\partial_x\theta_t$. Consequently, the stability condition is very similar to plate rotation, albeit with a different rotation amount at each point. Like in the first example, eq.~\eqref{Bforcebalance} leaves us with gauge freedom to split the pattern between the top and bottom plates. This freedom allows us to fulfill the stability criterion at each point by choosing top-pattern gradients as necessary. Therefore, stability poses no real constraint on $B$.
        Nonetheless, because of non-locality we need to make sure that a point on the moving surface is not ``double-booked'', requiring that $\tfrac{\partial B}{\partial x}>1$ everywhere, namely the disclination line must travel slower than the top plate. 
        Notably, if $B$ is \emph{one-sided Lipschitz} with respect to $x$, then we may multiply it by a constant factor to fulfill this criterion, namely ask that the top plate is moved ``faster''.
        \end{itemize}
    }

    \begin{table}[ht]
        \centering
        \add{
        \resizebox{\textwidth}{!}{%
        \begin{tabular}{|c|c|c|}    
             \hline
             Parameter       
             & Surface patterns $\theta _b,\theta _t$ 
             & Stability criterion  
             \\ \hline
             Temperature     
             & $\overline{\Delta\theta} = -\frac{\sigma}{2\pi q} B\frac{B_{x}^{2}B_{yy}-2B_{x}B_{y}B_{xy}+B_{y}^{2}B_{xx}}              {(B_{x}^{2}+B_{y}^{2})^{3/2}}$                                        
            & $B_{y}^{2}B_{{xx}}-2B_{x}B_{y}B_{{xy}}+B_{x}^{2}B_{{yy}}<0$                          
            \\ \hline
            \makecell{Top plate \\ Rotation}
            & \makecell{$\theta_b = B+\frac{\sigma \lambda}{2\pi                             q}\frac{B_{x}^{2}B_{yy}-2B_{x}B_{y}B_{xy}+B_{y}^{2}B_{xx}}{(B_{x}^{2}+B_{y}^{2})^{3/2}},$\\
            $\theta_t=\pi/2$}
            & $q\hat{\boldsymbol{T}}_\perp\cdot\nabla B > 0$          
            \\ \hline
            \makecell{Top plate \\ Translation}
            &    \makecell{                
                 $\theta_{t}(x,y) =\chi(x+X^*,y)+\frac{1}{2}(q \pi-\frac{\lambda}{2\pi q}\kappa(x+X^*,y)),$\\
                 $\theta_{b}(x,y) =\chi(x,y)-\frac{1}{2}(q \pi-\frac{\lambda}{2\pi q}\kappa(x,y))$ \hspace{8ex} $\S$  }                    
         &
         $B(x,y)$ one-sided Lipschitz w.r.t. $x$ 
         \\ \hline      
        \end{tabular}
        }
        \caption{Explicit surface pattern formulae for realizing arbitrary curve families utilizing different control parameters. Curve families, represented by the input function $B(x,y)$, must satisfy the matching stability criterion for the desired disclination paths to be in stable equilibrium. $\S$ $\kappa(x,y)$ is defined in eq.~\eqref{kappa}; $X^*(x,y)$ is determined by the implicit equation $B(x+X^*(x,y),y)=X^*(x,y)$; $\chi(x,y)$ is an arbitrary gauge function.}
        \label{tab:my_label}}
    \end{table}

\section*{Discussion}
We have shown that surface patterning in the parallel-plane setup practically allows full control over the shape of disclination lines. Whenever an additional control parameter exists in the system, e.g. the temperature, relative translation or rotation between the boundary plates etc., a single surface pattern gives rise to an entire family of curves, that emerge as equilibrium disclination curves at different values of the control parameter, as illustrated in Figure~\ref{fig2}. For several realizable control parameters, we gave explicit formulae for the surface patterns as function of the desired family of curves. As shown, restrictions may apply on the families of curves that could be stably designed using our method as result of the local dependence on the control parameter. Nonetheless, in an experimental realization one typically has some flexibility over which control parameter to use; in which direction and how fast to change it to switch between the desired target curves, and an additional inherent gauge freedom in splitting the pattern between the top and bottom plate. This flexibility removes many of the restrictions and makes our protocol useful and handy for practical applications. Among these applications are optical devices, dynamical circuiting and printing devices (through disclination-based directed assembly \cite{blanc_ordering_2013,guo2024directing,wang_topological_2016}) and more.

\begin{figure}[ht]
\centerline{\includegraphics[width=1\columnwidth]{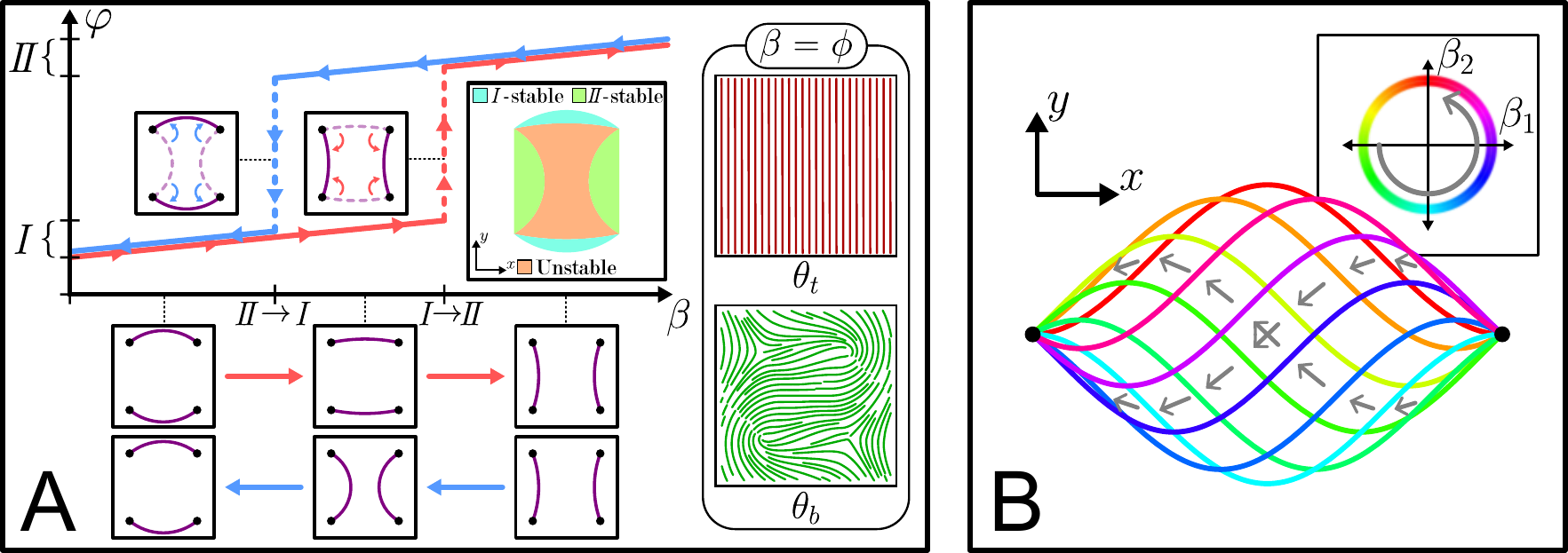}}
\caption{Applications. \add{(A)}\remove{Left:} Switch with controllable hysteresis. A curve-dependent parameter $\varphi$ (e.g. the angle between curve tangent and x axis at bottom left corner) is plotted as function of the control parameter $\beta$. Locally stable curves are found at two disconnected regimes, $I$ or $I\!I$, corresponding to distinct connectivity states. It is possible for $I$ and $I\!I$ to be simultaneously locally stable in a range of $\beta$ values, resulting in hysteresis; the connectivity profile is different for $\beta$-increasing (red) and $\beta$-decreasing (blue) processes. \add{Center inset}\remove{Inset}: the limits of stability for each of the two regions is set by the chosen surface patterns. The values for the $I\to I\!I$ and $I\!I\to I$ transitions can thus be tuned independently, allowing full control over the hysteresis loop. \add{Left inset: Surface patterns (Dataset S5) that realize the hysteresis switch through the use of plate rotation ($\beta=\phi$). With critical angles $\frac{\pi}{8},0$ corresponding to the transitions $I\to I\!I$ and $I\!I\to I$ respectively.} \add{(B)}\remove{Right:} Continuous nontrivial cycle in shape space. Each set of control parameter values $\beta_1, \beta_2$ (inset) corresponds to a single curve (of matching color) in real space. A continuous change of parameters results in a continuous change of curve shape. A nontrivial loop in parameter space, here a CCW cycle, gives rise to a nontrivial periodic cycle in shape space, here a left-moving wave, as indicated by the gray arrows.}
\label{fig4}
\end{figure}

In the above analysis we only discussed curve homotopies for the sake of obtaining explicit formulae, however this it is not generally required that paths change continuously. Multiple connectivity alternatives in systems with more than a pair of surface defects commonly give rise to multistability, abrupt path changes, and hysteresis \cite{frankread,nys_nematic_2022,jiang2022active,harkai_electric_2020,guo_photopatterned_2021}. The work that we have presented here allows full characterization and control over the stability of each branch at each value of the control parameter. We may therefore fine-tune the loss of stability of each branch at a desired value of the control parameter, and thus accurately design hysteresis loops (Fig.~\ref{fig4}A). This feature is key for possible applications like smart switches, actuation devices, or parameter-sensitive gauges.

Moreover, it is reasonable to consider an experimental system in which \emph{two or more} system parameters are controlled, e.g., the rotation between plates \emph{and} the temperature. We are not aware of reports of such a system in the literature, but this is likely only because such setup was not considered useful in the past. As an immediate first application, such control will effectively remove stability restrictions; a local shape-change that cannot be achieved by heating can be achieved by rotation and vice versa, and we may pick a non-trivial trajectory in parameter space to achieve the desired trajectory in shape space. Furthermore, it will be possible to realize nontrivial cycle in the set of shapes, not through hysteresis but rather by nontrivial closed loop in the set of parameters (Fig.~\ref{fig4}B). \add{For example, one may create an apparatus that pushes immersed colloids to the left when heated, raised, cooled and then lowered, iteratively. If the cycle is reversed, colloids will be pushed to the right.}
Other applications may make use of the full dimensionality of parameter space to realize within the same system larger (namely two-parameter) families of curves.

\section*{Acknowledgments}
The authors thank Jonathan Selinger and Charles Rosenblatt for fruitful discussions. This research was supported by the Israel Science Foundation (grant No. 2677/20).

\clearpage          
\pagenumbering{roman}  
\setcounter{page}{1}   
\newcounter{datasetcount}

\newcommand{\dataset}[2]{%
  \stepcounter{datasetcount}%
  \textbf{Dataset \arabic{datasetcount}:} (#1) - #2 \\%
}

\begin{center}
    \section*{Supporting Information\\Inverse Design of Parameter-Controlled Disclination Paths}
\end{center}

\author{}
\date{}

\subsection*{Appendix A -- Effective Energy Functional\textbackslash Boundary terms}

We can construct the energy functional of this theory to further understand
energy contributions of the boundary terms, we take the energy functional
to be
\[
U\{\Gamma\}=\frac{K h}{2}\int_{\mathbb{R}^{2}}\tau^{2}ds+\gamma\int_{\Gamma}dl=\frac{K}{2h}\int_{\mathbb{R}^{2}}(\overline{\Delta\theta}+2\pi qm_{\Gamma})^{2}ds+\gamma\int_{\Gamma}dl
\]

where we have replaced $h\,\tau=\overline{\Delta\theta}+2\pi q m_{\Gamma}$ for the total twist such that, across $\Gamma$, $\overline{\Delta\theta}\equiv \tfrac{h}{2}\left(\left.\tau\right|_{\Gamma^+}+\left.\tau\right|_{\Gamma^-}\right)$ is continuous and $m_{\Gamma}\in\tfrac{1}{2}+\mathbb{Z}$ jumps from $-\tfrac{1}{2}$ to $\tfrac{1}{2}$. The twist $\tau$ is continuous on $\mathbb{R}^2\setminus\Gamma$. On the other hand, $\overline{\Delta\theta}$ and $m_\Gamma$ may have discontinuities elsewhere, however these do not affect the functional derivative with respect to $\Gamma$ (see below).

The functional derivative with respect to the line defect position is
then

\[
\delta U=\int_{\Gamma}\left(\frac{2\pi qK}{h}\overline{\Delta\theta}+\gamma\kappa_{s}\right)\delta\Gamma_{\perp}dl+\gamma\left.\delta\Gamma_{\perp}\cdot\boldsymbol{n}_{\Omega}\right|_{\text{endpoints}}.
\]

The original force equation is recovered along with boundary terms
that account for changes in length for a perturbed curve with different
endpoints. In finite domains where defect lines intersect the boundary,
the boundary term dictates that the defect is perpendicular to the boundary
for stability.

\add{
    \newpage
    \subsection*{Appendix B -- Stability analysis in the inverse design framework and explicit pattern formulae}

    \paragraph{Stability.}
    Taking the differential of eq.~\eqref{forcebalance} with respect to $\beta$, we obtain
    
    \begin{equation}
        0= \frac{d f}{d\beta} = \frac{\partial f}{\partial \beta} +
        \frac{\delta f}{\delta \Gamma_\beta}\frac{\partial \Gamma_\beta}{\partial \beta}.
    \end{equation}
    Multiplying by the dot product $\hat{\boldsymbol{T}}_\perp\cdot\nabla B$ we get
    \begin{equation}
        \begin{split}
            0&= \frac{\partial f}{\partial \beta}(\hat{\boldsymbol{T}}_\perp\cdot\nabla B) +
            \frac{\delta f}{\delta \Gamma_\beta}\frac{\partial \Gamma_\beta}{\partial \beta}(\hat{\boldsymbol{T}}_\perp\cdot\nabla B)\\
            &= \frac{\partial f}{\partial \beta}(\hat{\boldsymbol{T}}_\perp\cdot\nabla B) +
            \frac{\delta f}{\delta \Gamma_\beta}(\frac{\partial\boldsymbol{\Gamma}_\beta}{\partial\beta}\cdot\nabla B)\\
            &=\frac{\partial f}{\partial \beta}(\hat{\boldsymbol{T}}_\perp\cdot\nabla B) +
            \frac{\delta f}{\delta \Gamma_\beta},
        \end{split}
    \end{equation}
    where we have used the relation $\frac{\partial \boldsymbol{\Gamma}_\beta}{\partial \beta}\cdot\nabla B = 1$ that is implied by the definition \eqref{Bdef}. Using the stability criterion \eqref{stableforce}, we get
    \begin{equation}\label{stabilityB}
            \frac{\partial f}{\partial \beta}(\hat{\boldsymbol{T}}_\perp\cdot\nabla B) =
        -\frac{\delta f}{\delta \Gamma_\beta}>0.
    \end{equation}
    \newpage
    \subsection*{Appendix C -- Explicit pattern formulae and stability criteria}
    
    \subsubsection*{I. Temperature}
        We choose the control parameter $\beta$ to be the smoothing scale $\lambda$. In this case, the force $f=\lambda\kappa+\widetilde{\Delta\theta}$ with $\beta=\lambda$ renders eq.~\eqref{Bforcebalance} in the form
        \begin{equation}\label{theta1}
            \overline{\Delta\theta} = -\frac{1}{2\pi q}B\kappa(x,y) = -\frac{\sigma}{2\pi q} B\frac{B_{x}^{2}B_{yy}-2B_{x}B_{y}B_{xy}+B_{y}^{2}B_{xx}}        {(B_{x}^{2}+B_{y}^{2})^{3/2}}.
        \end{equation}
        
        \subparagraph{Stability.}
        From $f=\lambda\kappa+\widetilde{\Delta\theta}$ we get $\tfrac{\partial f}{\partial \lambda}=\kappa$, which turns          eq.~\eqref{stabilityB} into
        \begin{equation}\label{Stability1}
        \boldsymbol{\kappa}\cdot\nabla B = \kappa\hat{\boldsymbol{T}}_\perp\cdot\nabla B > 0.
        \end{equation}
        Thus, stability requires that an increase in $\lambda$ must induce positive curvature flow, namely locally shorten the curve at every point. We can rewrite this condition only using $B$:
        \begin{equation}\label{Stability1B}
        B_{y}^{2}B_{{xx}}-2B_{x}B_{y}B_{{xy}}+B_{x}^{2}B_{{yy}}<0.
        \end{equation}
        Given a set of input curves $B$, eq.~\eqref{Stability1B} is a quick sorting criterion to determine whether $B$ is realizable via temperature change, and if so in which direction.

        \subsubsection*{II. Plate rotation}
        The control parameter $\beta = \phi$ in this case is the top plate rotation angle, thus $\widetilde{\Delta\theta} = 2\pi q(\phi-\theta_b)$. Then, from eq.~\eqref{Bforcebalance}:
        \begin{equation}\label{theta2}
        \theta_b = B+\frac{\lambda\kappa(x,y)}{2\pi q} = B+\frac{\sigma \lambda}{2\pi q}\frac{B_{x}^{2}B_{yy}-2B_{x}B_{y}B_{xy}+B_{y}^{2}B_{xx}}{(B_{x}^{2}+B_{y}^{2})^{3/2}}
        \end{equation}

        \subparagraph{Stability.}
        From $f=\lambda\kappa+\widetilde{\Delta\theta}$ we get $\tfrac{\partial f}{\partial \phi}=2\pi q$, which turns eq.~\eqref{stabilityB} into
        \begin{equation}\label{Stability2}
        2\pi q\hat{\boldsymbol{T}}_\perp\cdot\nabla B > 0.
        \end{equation}
        Stability implies that a curve can deform only towards one of its sides as the top plate is rotated. This direction is $q\hat{\boldsymbol{T}}_\perp$ for CCW rotation and $-q\hat{\boldsymbol{T}}_\perp$ for CW rotation. Equivalently, at its ends the curve must co-rotate with the top plate around a positively charged surface defect on the bottom plate, and counter-rotate with the top plate around a negatively charged defect.

        \subsubsection*{III. Plate translation}
        We consider a translation of the top plate in some direction, say $\hat x$, by distance $\beta = X$. We then get $\widetilde{\Delta\theta}=2\pi q(\theta_t(x-X,y)-\theta_b(x,y)\pm q\pi)$. We may write the top and bottom plate patterns using a gauge function $\chi(x,y)$ in the form
        \begin{equation}\label{theta3}
        \begin{split}
        \theta_{t}(x,y) & =\chi(x+X^*(x,y),y)+\frac{1}{2}(q \pi-\frac{\lambda}{2\pi q}\kappa(x+X^*(x,y),y)),\\
        \theta_{b}(x,y) & =\chi(x,y)-\frac{1}{2}(q \pi-\frac{\lambda}{2\pi q}\kappa(x,y)),
        \end{split}
        \end{equation}
        where $X^*(x,y)$ is determined by the implicit equation
        \begin{equation}\label{Xstar}
            B(x+X^*(x,y),y)=X^*(x,y).
        \end{equation}
        Solving for $\theta_{t}(x,y)$ requires first solving eq.~\eqref{Xstar} and then substituting it into eq.~\eqref{theta3}.

        To obtain a single-valued function for $\theta_{t}$, one must demand a unique solution to eq.~\eqref{Xstar}. Thus, we require
        that $\tfrac{\partial B}{\partial x}>1$ everywhere, namely the disclination line must travel slower than the top plate. 
        Notably, if $B$ is \emph{one-sided Lipschitz} with respect to $x$ then we may multiply it by a constant factor to fulfill this criterion, namely ask that the top plate is moved ``faster''.

        \subparagraph{Stability.}
        From eq.~\eqref{simpleforce} we get $\tfrac{\partial f}{\partial X}=-2\pi q\tfrac{\partial \theta_t}{\partial X}\bigl|_{\boldsymbol{\Gamma}_{X}-X\boldsymbol{\hat{x}}}$, which turns eq.~\eqref{stabilityB} into
        \begin{equation}\label{Stability3}
        2\pi \frac{\partial \theta_t}{\partial X}\biggl|_{\boldsymbol{\Gamma}_{X}-X\boldsymbol{\hat{x}}}q\hat{\boldsymbol{T}}_\perp\cdot\nabla B <0.
        \end{equation}
       We can use our gauge freedom in $\chi$ to fulfill this condition, thus stability poses no real constraint on $B$.     

       \newpage
        \subsection*{Datasets} Surface patterns that appear in figures 2-4 in CSV format. The values in each file represent $\theta_t$ or $\theta_b$ measuring the director angle w.r.t. the x-axis in radians, as function of x (rows) and y (columns).\\~\\
        
       \dataset{theta\_b\_phi\_N.csv}{Bottom plate pattern $\theta_b$, for realizing the cat curves using top plate rotation, with a constant top pattern taken to be $\theta_t=\frac{\pi}{2}$ (Fig.~2 and Fig.~3 [middle row]).}
       
       \dataset{theta\_b\_lambda\_N.csv}{Bottom plate pattern $\theta_b$, for realizing the cat curves using temperature, with a constant top pattern taken to be $\theta_t=\frac{\pi}{2}$ (Fig.~3 [top row]).}
       
       \dataset{theta\_b\_X\_N.csv}{Bottom plate pattern $\theta_b$, for realizing the cat curves using top plate translation (Fig.~3 [bottom row]).}
       
       \dataset{theta\_t\_X\_N.csv}{Top plate pattern $\theta_t$, for realizing the cat curves using top plate translation (Fig.~3 [bottom row]).}
       
       \dataset{theta\_b\_phi\_switch.csv}{Bottom plate pattern $\theta_b$, for realizing the hysteresis switch, with a constant top pattern taken to be $\theta_t=\frac{\pi}{2}$ (Fig.~4 [A]).}
  }


\end{document}